\documentclass[twocolumn,twoside,slac_two]{revtex4}
\usepackage{graphicx}
\usepackage{fancyhdr}
\begin{document}

\title{Using the Big Ideas in Cosmology to Teach College Students}

%

\author{McLin, K. M., Cominsky, L. R., Metevier, A. J.}
\affiliation{Sonoma State University, 1801 East Cotati Ave, Rohnert Park, CA 94928, USA}
\author{Coble, K.}
\affiliation{Chicago State University, Department of Chemistry and Physics, 9501 South King Dr., Chicago, IL, 60628, USA}
\author{Bailey, J. M.}
\affiliation{University of Nevada, Las Vegas, Department of Teaching and Learning, 4505 South Maryland Parkway, Box 45305., Las Vegas, NV, 89154, USA}

\begin{abstract}
Recent advances in our understanding of the Universe have revolutionized our view of its structure, composition and evolution. However, these new ideas have not necessarily been used to improve the teaching of introductory astronomy students. In this project, we have conducted research into student understanding of cosmological ideas so as to develop effective web-based tools to teach basic concepts important to modern cosmology. The tools are intended for use at the introductory college level. Our research uses several instruments, including open-ended and multiple choice surveys conducted at multiple institutions, as well as interviews and course artifacts at one institution, to ascertain what students know regarding modern cosmological ideas, what common misunderstandings and misconceptions they entertain, and what sorts of materials can most effectively overcome studentsÕ difficulty in learning this material. These data are being used to create a suite of interactive, web-based tutorials that address the major ideas in cosmology. One common  misconception that students in our introductory courses possess is that scientific explanations are ``made up,'' and not supported by observational data. Having students engage with real data is a powerful means to help students overcome this misconception. For this reason, the tutorials we are developing include authentic student interaction with actual data where possible. Students master the scientific concepts and reasoning processes that lead to our current understanding of the universe through interactive tasks, prediction and reflection, experimentation, and model building. This workshop will demonstrate the use of some of the modules we have created and will allow participants to test the modules for themselves.

\end{abstract}

\maketitle

\thispagestyle{fancy}


\section{Introduction}
The past several decades have seen a revolution in our understanding of cosmology. Cosmological parameters like the Hubble constant have been measured to unprecedented precision \citep{freedman:2001}, as has the geometry of the universe \citep{lange:2001,spergel:2003}. In addition, observations of distant supernovae have shown that the cosmic expansion rate is increasing with time \citep{riess:1998,perlmutter:1999}, and surveys of galaxies have revealed how cosmic structures have evolved over time \citep{abazajian:2003}. These observations and others, along with powerful computer models, have given us unprecedented insight into how the universe began and how it has evolved and continues to evolve. However, these new ideas have not necessarily been used to improve our teaching of introductory astronomy students. The disconnect is partly caused by the fast pace at which cosmological discoveries have accumulated; textbook publishers and course instructors have been hard-pressed to keep up. A contributing factor has been the low emphasis given to cosmological topics in typical introductory textbooks \citep{bruning:2006a,bruning:2006b} and placement of cosmology in the last week or two of the typical introductory course. The lack of cosmological focus is unfortunate  given what we now about the origin and evolution of the Universe. What's more, cosmological topics are often of great interest to students \citep{pasachoff:2002}.  

The Big Ideas in Cosmology project puts recent cosmological advances at the fore, making them the primary focus of a 15 week long course for general education (GE) undergraduate students. The underlying ideas behind the curriculum design are that interaction with data is vital if students are going to fully understand the material, and that preexisting concepts held by the students  can inhibit their learning and must be addressed directly. The curriculum employs computer interactive simulations that allow students to ``measure" spectra, brightness, and other relevant properties of systems being studied, and then allow for the students to formulate or test models to explain the data. In addition, the interactives and text materials are designed to address concepts with which research shows students often have difficulty. 

This article describes the research and curriculum development project, as well as several of the interactive materials that have been developed thus far. The purpose of our workshop at the ASP was to have participants learn about the curriculum and try out some of the interactives. The remainder of the article will describe the basic organization of the cosmology curriculum, give an overview of the research program behind it, and an idea of the remaining development and evaluation methods we will use to complete it.

\section{The Curriculum}

The cosmology curriculum is being designed to address student misconceptions whenever possible. Recent  publications from \citet*{wallace:2012d} and our own group \citep{bailey:2012,coble:2012,trouille:2012} have documented some of these student misconceptions. The curriculum includes traditional text, static imagery and video, as well as computer-based interactive activities. Emphasis is given to student experiences with real datasets and measurements of the data. Because most cosmological models are mathematical in nature, we have introduced simple math exercises that use graphing and proportions throughout, moving from simpler to more complex treatments as the course progresses. In addition, students are asked to reflect upon the material they are learning. They must give feedback to the computer learning system so that both they and their instructor have a running tally of their progress through the material.

\subsection{Organization}

The course is organized into three thematic modules, and each module is divided into five chapters. A chapter covers approximately one week in an introductory course.

The first module is focused on basic ideas of size and time scales of the Universe and its contents. The purpose of this module is to bring students to a level of comfort when thinking about relevant size scales, since research and experience show that they often have difficulty doing so. For instance, our own research shows that students can have difficulty differentiating between objects like the solar system, the galaxy and the Universe, and between the relative sizes of each. Module One also includes a chapter on the timescales relevant to studying cosmology. Topics include the history of Earth and the solar system, timescales of stars and galaxies, etc. There is also a chapter in which light and telescopes are explored - since this curriculum is intended for an introductory course, we do not assume that the students will have learned about the electromagnetic spectrum previously. Finally, the last chapter of the module introduces the modifications of space and time inherent to special relativity, including spacetime and invariance of the laws of physics. This chapter is useful later when students learn about general relativity, but much of it can be skipped if instructors wish.

Module Two explores gravitation from a  Newtonian framework and from the curved spacetime perspective of general relativity. This material is then applied to learning about dark matter and the various lines of evidence for it. The module introduces some of the more involved treatments of data analysis and modeling, albeit at a simple level. Module Two also includes a chapter on black holes, as they are of general interest to students, and they provide a good application of ideas from general relativity. The treatment of relativity, both special and general, goes beyond what is typical for an introductory course. We try to avoid the often confusing visuals used to discuss spacetime and its curvature, and instead use a description more focused on how curvature can be described by the effect it has on distances between points in a space. 

Module Three addresses most of the fundamental concepts of cosmology. These include cosmic expansion, conditions in the early Universe, the cosmic microwave background and large scale structure, and dark energy and the fate of the Universe. Each of these topics is given its own chapter so that it can be explored in some detail. In addition, many of the tools developed in earlier modules are finally brought to full flower in Module Three. Ideas from Newtonian and Einsteinian gravity, electromagnetic radiation, distance measurement techniques, and others, are employed to understand the modern scientific conception of how the Universe has been evolving since its formation. In addition, we continue to use and build upon mathematical tools like graphing and model fitting that have been employed in previous chapters. 

An overview of the three modules and the chapters they contain can be seen in Table \ref{tab:curriculum}.

%


\begin{table*}[!ht]
\caption{Organization of the Curriculum}
\smallskip
\begin{center}
   \begin{tabular}{ll} 
	\tableline
	\noalign{\smallskip}
      Module    & Chapter\\
	\tableline
	\noalign{\smallskip}
      1. Our Place in the Universe:       & 1. The Size and Scope of Space \\
              Space and Time  			& 2. Observing the Universe: Light and Telescopes   \\
                						& 3. Motion and Time   \\
                 						& 4. Measuring Distances   \\
                              				& 5. Special Relativity   \\
	\tableline			
      2. The Dark Side of      			& 6. Classical Physics: Gravity and Energy  \\
	 the Universe: Gravity,      		& 7. Dark Matter  \\
	 Black Holes and      			& 8. General Relativity  \\
	 Dark Matter       			& 9. Black Holes  \\
              						& 10. Gravitational Lensing  \\
	\tableline					
      3. Our Evolving Universe:      	& 11. Cosmic Expansion and the Hubble Flow  \\
	Past, Present  and    			& 12. The Early Universe  \\
	Future                          		& 13. The CMB and Large Scale Structure  \\
	                              			& 14. Dark Energy  \\
              						& 15. Cosmic Concordance and the Fate of the Universe  \\
  \noalign{\smallskip}						
  \tableline
   \end{tabular}
   \end{center}
   \label{tab:curriculum}
\end{table*}


\subsection{Pedagogical Approach}

\begin{figure*}[!ht]
\centering
\includegraphics[width=135mm]{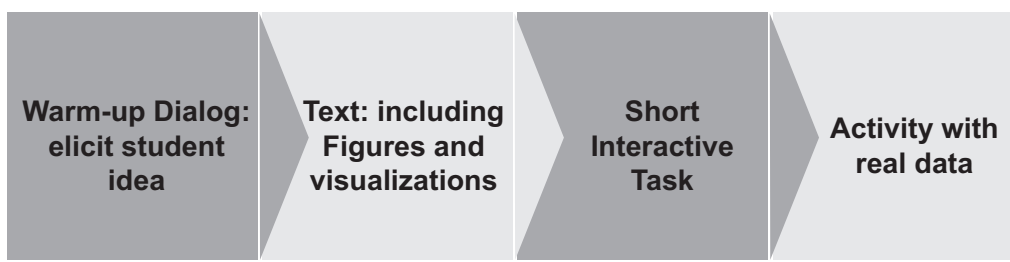}
\label{fig:pedflow}
\caption{Flow chart of the pedagogical design for each section. Chapters typically contain several sections, each of which explores a single topic related to the subject of the chapter.}
\end{figure*}

Our goal is to approach this material in a more sophisticated manner than is generally found in the typical text. We chose an online format to enable the use of interactive activities, which are integral to student learning. The curriculum does include plain text and static images, but we emphasize that the computer exercises are not ``supplementary.'' They are a vital part of the material and are built right into the flow of the text. 

Each chapter has the same format. It begins with a short video intended to introduce the primary theme of the chapter and get students thinking about it. The chapter is then divided into sections that cover one particular topic, and these sections are introduced using fictional ``dialogues" between archetypal students.  When possible, the dialogues use student-derived language to discuss the topic of that particular section, with phrases gleaned from our research. The dialogues address commonly held misconceptions when appropriate. Our intent is to introduce new science topics in the students' own language, and to address misconceptions that we and others have found are common to the target audience of GE students. Research has shown that students learn better when new concepts are based upon their existing ideas about the world \citep{bransford:1999}, and so we incorporate their ideas from the outset.  The students are asked to think about each dialogue and choose one of the fictional speakers with whom they most agree and explain why. A sample dialogue is given below:

\begin{quote}
Some students are looking through a telescope at a star.

Alicia says, ``How many light years away do you think this star is? Do you think we could measure the distance to it?''

Bill replies, ``No way; itÕs physically impossible because we can't go there.''

``I donÕt think we have to go there if we use the correct method,'' Corey says, ``like in geometry class when we learned how to triangulate distances on construction sites.''

``But I thought that would only work on Earth because we are there,'' Alicia says.

Corey says, ``I think it will work everywhere; itÕs just a different scale.''

Which student(s) do you agree with, and why?

a. Alicia\\
b. Bill\\
c. Corey\\
d. None\\
\end{quote}

The dialogues are followed by a text passage, with illustrations or visualizations as needed, and then an exercise of some sort. The exercises can be ranking tasks, simple numerical exercises or a short interactive computer activity. Throughout the lesson students are asked to record their thoughts and answers to questions for later review. Students are given automatic feedback about the acceptability of their answers, and their responses are recorded so that they and the instructor have a running record of how their thinking evolves over time. 

Each chapter will generally have several sections as described above. When all the important topics for a chapter have been explored, the chapter closes with a longer summary ``Wrapping it Up'' activity, and an exercise called a ``Mission Report.'' The Wrapping It Up generally involves the students working with actual datasets and integrating the important topics in the chapter. The Mission Report is graded for accuracy and is an easy way for the instructor to assess whether or not the student has mastered the most important learning objectives in the chapter. It also provides  an opportunity for the student to provide feedback about the learning efficacy of the chapter materials. A schematic overview of the pedagogical flow is shown in Figure 1.

%



\subsection{Sample Interactives}

An integral part of each chapter is a set of computer interactive exercises. In this section we describe one of these in order to give a flavor for how they are used in the curriculum. The example described is part of Chapter 4, on the cosmic distance scale. We cite the main sequence (MS)  fitting method, which is a standard candle distance method. In this method, the MS of a cluster is compared to a fiducial one with known distance. The offset in brightness on a color-magnitude diagram between the observed cluster and the fiducial gives their relative distance. The computer exercise allows the student to slide the observed cluster's MS up and down until it overlays the fiducial. The interactive tool gives a running display  of the offset in magnitudes and converts it to a distance for the observed cluster. No attempt is made to get a ``best fit'' in this exercise, nor do we worry about complications like extinction. The goal is that students will understand that the difference in brightness between objects is related to their distance from us. Similar interactive exercises are used to teach the other rungs on the cosmic distance ladder. A detail from the MS fitting exercise is shown in Figure 2.   

\begin{figure}[!ht]
\centering
\includegraphics[width=75mm]{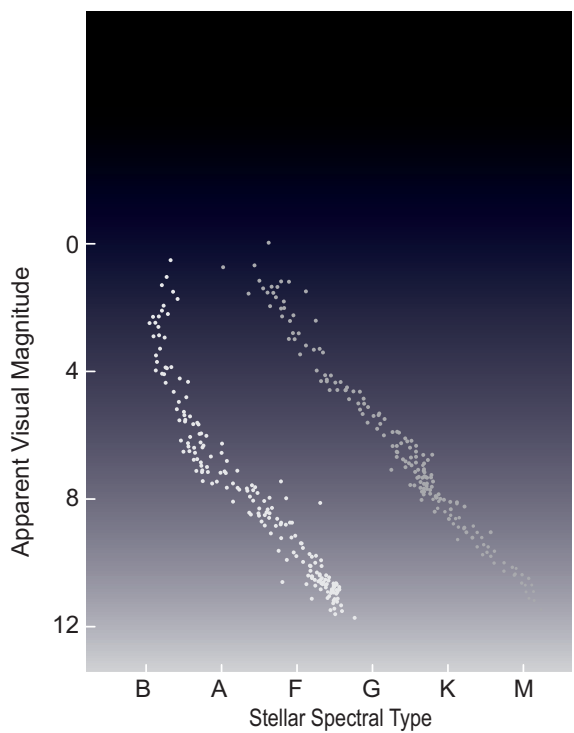}
\label{fig:msfit}
\caption{Detail of the main sequence fitting tool that allows students to slide the MS of one star cluster up and down until it overlies that of a fiducial cluster of known distance.}
\end{figure}

\section{Research Overview}

This curriculum is strongly based upon results from research into the understanding that undergraduate students bring to their GE astronomy classes. We have conducted open-ended  surveys at multiple institutions as well as in-depth studies at a single institution. The research is ongoing, and we have recently published the first of our results. The first publication \citep{bailey:2012} describes the findings from a set of open-ended questionnaires that explore student understanding of the size, structure, composition and evolution of the Universe. These surveys were done prior to instruction, with a total of 1270 students participating; more than 200 responded to any of three surveys. The research uncovered many areas of difficulty for students. For instance, more than a quarter of students had difficulty explaining the differences between the solar system, the Galaxy and the Universe.  They also were unsure of the hierarchal relationships between them. Fewer than a third of the students thought that the age of the Universe was measured in billions of years, and more than a quarter did not even attempt an answer for this item. In addition, some students did not even believe that there was {\it or even could be} scientific evidence for some of the subjects queried; fewer than 20\% thought that it was possible to know anything about the age of the Universe. One important finding, consistent with earlier studies, was that many students have trouble with the specialized language of science. For instance, terms like light-year or Big Bang can be confusing when students attempt to interpret their meaning in the context of everyday usage. Care must be taken by instructors to ensure that their students appreciate the special meanings these terms have in a scientific context. 

Another two papers are forthcoming:  \citep{coble:2012,trouille:2012}. The first of these focuses on distance and size scales. The second  looks at student understanding of relevant timescales in cosmology and the Big Bang Theory. These studies are based on results covering several semesters of an introductory course in astronomy at an urban minority-serving institution. Research instruments included: essays given as homework, open-ended surveys given after lecture but before lab, exam questions (essay, MC, TF, FIB) and interviews. Data was collected over five semesters and spans the entire course, from pre-instruction through the final exam. Depending on the instrument and question, N $\sim$60 for courses artifacts. A subset of 15 students was interviewed in order to delve more deeply into their thoughts about cosmology. In general, the results from these studies corroborate those from \citet{bailey:2012}. Unsurprisingly, students do not know  a lot about cosmology at the start of the term. However, student understanding can increase through the semester and is greatly enhanced if the students are given relevant and engaging exercises; lecture alone was not seen to improve student understanding significantly. Students' poor math skills are a hinderance to mastering new cosmological concepts, but continued practice with activities that involve making and interpreting graphs, and working with proportional relationships, can improve these skills over the semester. Disturbingly, it is found that students have difficulty letting go of some of their preconceived ideas about the size of the Universe. Working with real data seems to help students understand the evidence behind scientific world views, and also to discard some of their mistaken preconceptions. Since basic math skills are needed for most data-driven activities, helping students attain these, especially in terms of working with graphical data and using simple proportional relationships to guide their reasoning, can pay large dividends over the course of the semester. 

Space here does not allow for more than a short synopsis of some of our research results. Readers who are interested in the details of these studies and their findings are urged to read the papers referenced.

\acknowledgements This three-year project is being funded by the Education and Public Outreach program for NASA's Fermi Gamma-ray Space Telescope, grant NNX-10AC89G from NASA's EPOESS program, and from the Illinois Space Grant Consortium. Background research was also funded in part by the National Science Foundation CCLI Grant \#0632563 at Chicago State University. We gratefully acknowledge the artistic and technical expertise of the Educational and Public Outreach Group at Sonoma State University, especially Aurore Simonnet and Kevin John, and the help of our current and former research students on the project: Melissa Nickerson, Carmen Camarillo, Virginia Hayes, Donna Larrieu, K'Maja Bell at Chicago State University; Roxanne Sanchez at UNLV; and Geraldine Cochran at Florida International University. We thank Ted Britton and his staff at WestEd for developing and analyzing the evaluation. Kendall-Hunt will be publishing the curriculum modules for national distribution. 

\bibliography{aspmclin}





\end{document}